\documentclass[]{iopart}

\usepackage[english]{babel}
\usepackage[utf8]{inputenc}

\expandafter\let\csname equation*\endcsname\relax
\expandafter\let\csname endequation*\endcsname\relax
\usepackage[]{amsmath,amssymb}
\usepackage[dvips]{graphicx}
\usepackage[breaklinks]{hyperref}
\usepackage[numbers,sort&compress]{natbib}
\usepackage[table]{xcolor}

\begin{document}

\title{Chiral Hadronic Mean Field Model including Quark Degrees of
  Freedom}

\author{P.\ Rau$^{1,2}$, J.\ Steinheimer$^{3}$, S.\ Schramm$^{1,2}$,
  H.\ St\"ocker$^{1,4}$}%

\address{$^1$Institut f\"ur Theoretische Physik, Goethe Universit\"at,
  Max-von-Laue-Str.\ 1, 60438 Frankfurt am Main, Germany}%
\address{$^2$Frankfurt Institute for Advanced Studies (FIAS),
  Ruth-Moufang-Str.\ 1, 60438 Frankfurt am Main, Germany}%
\address{$^3$Lawrence Berkeley National Laboratory,
  Berkeley, CA 94720, USA}
\address{$^4$GSI Helmholtzzentrum f\"ur Schwerionenforschung GmbH,
  Planckstr.\ 1, 64291 Darmstadt, Germany}%

\ead{rau@th.physik.uni-frankfurt.de}%

\begin{abstract}
  In an approach inspired by Polyakov loop extended NJL models, we
  present a nonlinear hadronic SU(3) $\sigma$--$\omega$ mean field
  model augmented by quark degrees of freedom. By introducing the
  effective Polyakov loop related scalar field $\Phi$ and an
  associated effective potential, the model includes all known
  hadronic degrees of freedom at low temperatures and densities as
  well as a quark phase at high temperatures and densities. Hadrons in
  the model exhibit a finite volume in order to suppress baryons at
  high $T$ and $\mu$ This ensures that the right asymptotic degrees of
  freedom are attained for the description of strongly interacting
  matter and allows to study the QCD phase diagram in a wide range of
  temperatures and chemical potentials. Therefore, with this model it
  is possible to study the phase transition of chiral restoration and
  deconfinement. In this paper, the impact of quarks on the resulting
  phase diagram is shown. The results from the chiral model are
  compared to recent data from lattice QCD.
\end{abstract}

\pacs{12.38.-t, 11.30.Rd, 14.20.Gk, 25.75.Nq}
\submitto{\JPG}

\maketitle
\section{Introduction}
\label{sec:introduction}

Deducing the properties of strongly interacting QCD matter at high
temperatures and densities is one of the major goals of heavy ion
collision experiments, as performed at the Relativistic Heavy Ion
Collider (RHIC), the Large Hadron Collider (LHC), or at the upcoming
Facility for Antiproton and Ion Research (FAIR). Experiments at the
LHC and RHIC strongly suggest that at sufficiently large beam
energies, i.e.\ high temperatures $T$ and small baryochemical
potentials $\mu_B$, a new state of matter is created, comparable to a
nearly perfect fluid with very low viscosity~\cite{Arsene2005,
  Back2005, Adams2005, Adcox2005}. Possible phase transitions in other
regions of the phase diagram with larger chemical potentials and
baryonic densities are probed at lower beam energies.\par
Quantum chromodynamics (QCD) is the theoretical framework for the
description of strongly interacting matter. A great difficulty in
dealing with the QCD Lagrangian results from the fact that the
equations of QCD cannot be solved analytically and perturbative
approaches on QCD are suitable only in the high-temperature regime.
Nevertheless, there is strong evidence for the existence of two phase
transitions for strongly interacting matter. One is the chiral phase
transition from a state with spontaneously broken chiral symmetry at
low temperatures to a chirally symmetric phase with vanishing
constituent quark masses~\cite{Nambu1961, Kirzhnits1972, Weinberg1974,
  Theis:1984qc}. For this transition, the chiral condensate $\sigma =
\langle \bar{q}q \rangle$ serves as a well-defined order parameter. At
vanishing baryochemical potentials $\mu = 0$, lattice QCD methods can
be used to study the chiral transition. In this theoretical approach,
the QCD Lagrangian is treated numerically on a space-time
grid. Calculations with different lattice actions consistently suggest
a smooth cross-over transition with a ``\emph{critical}'' temperature
in the region of $T_c \approx 160$~MeV ~\cite{Karsch2000, Fodor2002,
  Fodor2003, Fodor:2004nz, Karsch2005, Aoki2006, Cheng2006, Cheng2007,
  DeTar2008, Petreczky2009}. At non-zero chemical potentials, standard
lattice QCD methods are not applicable due to the fermion sign
problem. The fundamental phase structure of strongly interacting
matter is still unknown. However, expansion and reweighting methods
exist to extrapolate results from lattice QCD into the region $\mu >
0$ of the phase diagram~\cite{Fodor:2001pe, Fodor2002, Fodor2007,
  Allton:2002zi, Forcrand2003, D'Elia2003, D'Elia2007, Allton2005,
  Fodor2003, Forcrand2008, Kaczmarek2011, Wu2007}. In this region,
predictions on the order and exact location of the phase transition
from different lattice groups are not consistent and depend on the
used lattice action and other parameters such as the lattice spacing
and the quark mass. While often the existence of a first-order phase
transition and a critical endpoint at a certain chemical potential is
predicted~\cite{Stephanov1998, Stephanov1999, Fodor:2001pe,
  Fodor:2004nz, Stephanov2006}, other recent results suggest a smooth
cross-over transition in $\sigma$ for the whole $T$-$\mu$-plane under
investigation~\cite{Forcrand2008, Endrodi:2011gv}.\par
The other phase transition in QCD matter is the deconfinement
transition from hadronic bound quark states to a quark-gluon plasma
(QGP) state at high temperatures and baryonic
densities~\cite{Gyulassy2005} with the Polyakov loop $\Phi$ as a
suitable order parameter.\par
Besides direct numerical approaches solving the QCD Lagrangian, a
broad range of effective theoretical models exist for describing known
properties of the phase diagram and thermodynamic characteristics of
strongly interacting matter~\cite{Walecka1974, Serot1986, Boguta1977,
  Boguta:1981px, Khvorostukhin2007, Khvorostukhin2008, Nambu1961,
  Nambu1961b}. In this paper, such a model is presented including all
known hadronic degrees of freedom~\cite{Papazoglou:1997uw,
  Papazoglou:1998vr, Rau:2011av, Rau:2012px} as well as a quark phase
for the description of highly excited matter. With this model, it is
possible to study the QCD phase diagram with a particular focus on the
phase transitions. In the approach presented in the following, the
description of the hadronic phase is done by a mean field sigma-omega
type model and a quark phase is implemented following
Polyakov-loop-extended Nambu-Jona-Lasinio (PNJL)
models~\cite{Fukushima:2003fw, Fukushima:2008wg, Ratti:2005jh,
  Roessner:2006xn, Schaefer:2007pw, Sasaki:2006ww, Megias:2004hj,
  Ghosh:2006qh, Meisinger:1995ih, Steinheimer:2009hd,
  Steinheimer:2010ib}.\par
This paper is organized as follows. After this introduction, we
present the quark hadron model and outline its theoretical background
in Sec.~\ref{sec:model-theory}. Results for the order parameters of
chiral restoration and deconfinement phase transition and the
thermodynamic properties of the model presented in
Sec.~\ref{sec:results}. A short summary and concluding remarks are
given in Sec.~\ref{sec:summary}.

\section{Model approach}
\label{sec:model-theory}

For this study, we use a $SU(3)$-flavor $\sigma$-$\omega$ model with a
non-linear realization of chiral symmetry; see
Refs.~\cite{Boguta:1981px, Papazoglou:1997uw, Papazoglou:1998vr,
  Dexheimer:2008ax} for a more formal and detailed description of the
original purely hadronic model. In this effective model, the
Lagrangian in mean field approximation~\cite{Serot1986, Serot:1997xg}
takes the form
\begin{equation}
  \label{eq:lagrangian_summary}
  \mathcal{L} = \mathcal{L}_{\rm kin} + \mathcal{L}_{\rm int} +
  \mathcal{L}_{\rm meson},
\end{equation} 
where the first term represents the kinetic energy of the hadrons as
described in Ref.~\cite{Papazoglou:1998vr} in more detail. The
interaction of baryons and quarks with the scalar mesons fields
$\sigma$, $\zeta$ (attractive interaction, see
Eq.~(\ref{eq:effective_mass})) and the vector meson fields $\omega$,
$\phi$ (repulsive interaction) respectively, is expressed by
\begin{equation}
  \label{eq:L_int}
  \mathcal{L}_{int} = -\sum_i \bar{\psi_i} \left[  \gamma_0 \left( g_{i\omega}
      \omega^0 + g_{i\phi}
      \phi^0 \right) + m^*_i \right] \psi_i .
\end{equation}
The summation index $i$ runs over the three lightest quark flavors (u,
d, s), the baryonic octet, decuplet, and all heavier baryonic
resonance states up to masses of $m_{N^*} = 2.6$~GeV, which have at
least a three star rating in the the Particle Data Group
listings~\cite{Beringer:1900zz, Nakamura2010}.\par
The mesonic condensates in our model are considered in mean field
approximation. They include the scalar isoscalar field $\sigma \sim
\langle \bar{u}u + \bar{d}d \rangle$ (which is identified as the
chiral order parameter) and its strange equivalent $\zeta \sim \langle
\bar{s}s \rangle$, as well as the vector isoscalars $\omega$ and
$\phi$. The coupling strengths $g_{i \sigma, \zeta}$ of quarks and
baryons to the scalar fields account for their effective mass, which
to a large extent is generated dynamically
\begin{equation}
  \label{eq:effective_mass}
  m_{i}^* = g_{i\sigma}\sigma + g_{i\zeta}\zeta + \delta m_i .
\end{equation}
The explicit mass term $\delta m_i$ is chosen such as to reproduce the
tabulated vacuum masses, which is the mean mass for broad resonance
states. The lightest baryons exhibit an explicit mass of $\delta m_i =
150$~MeV and somewhat larger values with increasing vacuum mass of the
specific particle. The quark masses are set to $\delta m_{u, d} =
6$~MeV and $\delta m_s = 105$~MeV. According to
Eq.~\eqref{eq:effective_mass} the decrease of the $\sigma$-field at
high temperatures and densities leads to decreasing baryon masses and
thus to the restoration of chiral symmetry. The resulting effective
masses and the connected ground state properties are discussed in
detail at the beginning of Sec.~\ref{sec:results}.\par
In the same manner, the effective chemical potential for quarks and
baryons is given by
\begin{equation}
  \label{eq:effective_chem_pot}
\mu^*_i = \mu_i - g_{i \omega} \omega - g_{i \phi} \phi, 
\end{equation}
and therefore, its value is generated by the couplings $g_{i \omega,
  \phi}$ of quarks and baryons to the vector meson fields.\par
The mesonic part of the Lagrangian Eq.~\eqref{eq:lagrangian_summary}
\begin{equation}
  \label{eq:lagrangian_mesonic_part}
  \mathcal{L}_{\rm meson} = \mathcal{L}_{\rm vec} +\mathcal{L}_0 +
  \mathcal{L}_{\rm ESB}
\end{equation}
includes the mass terms and the self interactions of the vector mesons
\begin{align}
  \begin{split}
    \label{eq:self_int_vec_mes}
    \mathcal{L}_{\rm vec} &= \frac{1}{2} \frac{\chi}{\chi_0} \left(
      m^2_{\omega} \omega^2 + m^2_{\phi} \phi^2 \right) \\
    & \quad +g_4 \left( \omega^4 + \frac{\phi^4}{4} + 3 \omega^2
      \phi^2 + \frac{4 \omega^3 \phi}{\sqrt{2}} + \frac{2 \omega
        \phi^3}{\sqrt{2}} \right),
  \end{split}
\end{align}
as well as the self interaction terms of the scalar mesons explicitly
given by the terms
\begin{align}
  \begin{split}
    \label{eq:self_int_scalar_mes}
    \mathcal{L}_0 &= -\frac{1}{2} k_0' \, \left( \sigma^2 +
      \zeta^2 \right) + k_1 \, \left( \sigma^2 + \zeta^2 \right)^2 \\
    &\quad + k_2 \, \left( \frac{\sigma^4}{2} + \zeta^4 \right) + k_3' \,
     \sigma^2 \zeta \\
    & \quad- k_4 \, \chi^4 - \frac{1}{4} \chi^4\, \ln
    \frac{\chi^4}{\chi_0^4} + \frac{\delta}{3}\, \chi^4 \ln{
      \frac{\sigma^2 \zeta} {\sigma_0^2 \zeta_0} } ,
  \end{split}
\end{align}
and an explicit symmetry breaking, defined by the terms
\begin{align}
    \mathcal{L}_{\rm ESB} &= - \frac{\chi^2}{\chi_0^2} \left[ m_\pi^2
      f_\pi\sigma+\left(\sqrt{2}m_k^ 2f_k -\frac{1}{\sqrt{2}}m_\pi^ 2
        f_\pi\right)\zeta \right] .
\end{align}\par
The scalar dilaton field $\chi$ can be identified as the gluon
condensate. It was introduced in order to ensure scale invariance of
the Lagrangian and to model QCD trace
anomaly~\cite{Papazoglou:1998vr}. In the purely hadronic scenarios
without considering a quark phase, changes in the $\chi$-field are
very small and thus it is kept fixed at its ground state value
$\chi_0$. This does not apply, in the presence of the quark phase. In
this case, the quarks couple to the dilaton field [cf.\
Eq.~\eqref{eq:coupling-polchi}] and hence, its value changes with the
temperature. The chosen parameters of the Lagrangian are listed in
Tab.~\ref{tab:general-model-parameters}.
\begin{table}[b]
  \centering
  \begin{tabular}{|c|c|c|c|c|c|c|c|}
    \hline
    $k_0'$ & $k_1$ & $k_2$ & $k_3'$ & $k_4$ \\
    \hline
    $3.83 \times 10^5$~MeV$^2$ & 1.40 & -5.55 & $-1.07 \times
    10^3$~MeV & -0.23 \\
    \hline
  \end{tabular}
  \vspace{0.25em}\\
  \begin{tabular}{|c|c|c|}
    \hline
    $\delta$ & $\chi_0$ & $g_4$ \\
    \hline
    0.067 & 401.93~MeV & 38.5 \\
    \hline
  \end{tabular}
  \caption{General parameters used in  the mesonic part of the chiral
    model Lagrangian Eq.~\eqref{eq:lagrangian_mesonic_part}. The 
    couplings of the particles to the meson fields are given in 
    Tab.~\ref{tab:coupl-parameters}.} 
  \label{tab:general-model-parameters}
\end{table}\par
The mesonic potential, which sets the dynamically generated meson
masses, is defined as $\mathcal{V}_{\rm meson} = - \mathcal{L}_{\rm
  meson}$. From this, the masses of the mesons in the model are
obtained by the second derivative of the potential with respect to the
particular meson species $\xi_j$
\begin{align}
  \label{eq:effective_meson_mass}
  {m^*_j}^2 = \frac{\partial^2}{\partial \xi^2_j} \mathcal{V}_{\rm
    meson}.
\end{align}
\par
In our model the coupling strengths of the baryons and the quarks to
the fields are the parameters with the biggest impact on the resulting
mater properties and the phase diagram. The couplings of the baryonic
octet are fixed such as to reproduce the well-known vacuum masses and
nuclear saturation properties (see
Refs.~\cite{Dexheimer:2008ax,Dexheimer:2009hi} for details). For the
couplings of the baryonic resonances to the fields, a straight forward
two parameter ansatz is chosen~\cite{Rau:2011av}. This approach avoids
having multiple different coupling parameters for every single
resonance state or SU(3) multiplets. The resonance couplings are
connected to those of the nucleons to the respective fields via the
relations
\begin{align}
  \label{eq:couplings}
  g_{B_i \sigma, \zeta } &= r_{s} \; g_{N \sigma, \zeta},\\
  g_{B_i \omega, \phi}  &= r_{v} \; g_{N \omega, \phi},
\end{align}
where $r_s$ and $r_v$ are the so-called scaling parameters for the
scalar and vector fields. Ref.~\cite{Rau:2011av} demonstrates the
impact of baryonic resonance states on the phase diagram. This was
essentially done by varying the vector coupling strength (change the
value of $r_v$, which effectively controls the abundance of particles
at finite baryochemical potentials in the model). For reasonable
couplings, a first-order phase transition ceases to exist in favor of
a smooth cross-over due to the continuous population of heavier
resonances states.\par
In this study, the vector coupling parameter is kept fixed at $r_v =
0.9$. Furthermore, in order to ensure a smooth cross-over of the order
parameter and of the thermodynamic quantities at vanishing chemical
potentials, the scalar couplings are also fixed at a value close to
unity $r_s \approx 1$.  For the non-interacting hadron resonance gas
(HRG), the couplings of all particles to the fields are set to zero
and $\delta m_i$ is kept fixed at the respective vacuum mass
values~\cite{Beringer:1900zz, Nakamura2010}.\par
The couplings of the quarks to the meson fields are controlled by
$g_{q \sigma}$, $g_{q \zeta}$, $g_{q \omega}$, and $g_{q \phi}$, where
$q$ stands for the respective quark species (u, d, s). The explicit
masses of the quarks and the nucleons $\delta m_0$ and their coupling
strengths to the fields are listed in
Tab.~\ref{tab:coupl-parameters}. 
\begin{table}[bt]
  \centering
  \begin{tabular}{|c|c|cc|cc|cc|}
    \hline
    & $\delta m_0$ [MeV] & $g_\sigma$ & $g_\zeta$ & $g_\omega$ &
    $g_\phi$ \\
    \hline
    \hline
    u & 6.0   & -3.5 &  0.0 & 4.0 & 0.0\\
    d & 6.0   & -3.5 &  0.0 & 4.0 & 0.0\\
    s & 105.0 & 0.0  & -3.5 & 0.0 & 7.1\\
    \hline
    N & 150 & -9.83 &   1.22 & 11.56 & 0.0\\
    \hline
  \end{tabular}
  \caption{Explicit particle masses $\delta m_0$ and coupling
    strengths of the quarks and the nucleons $g_i$ to the fields. In
    the non-interacting HRG scenario, all couplings are set to zero
    (cf.\  Fig.~\ref{fig:T0_densities} for detailed parameter study).} 
  \label{tab:coupl-parameters}
\end{table}
In Sec.~\ref{sec:model_properties} a detailed study of the impact of
the quark couplings on the phase diagram and the nuclear ground state
is performed.\par
All thermodynamic quantities in our model can be derived from the
grand canonical potential, which is defined as
\begin{equation}
  \label{eq:grand_canon_pot}
  \frac{\Omega}{V} = -\mathcal{L}_{\rm int} - \mathcal{L}_{\rm meson}
  + \frac{\Omega_{\rm th}}{V}  - U_{\rm Pol}.
\end{equation}
In the heat bath of hadrons and quarks, the thermal contribution to
the potential includes those from quarks, baryons, and mesons
\begin{align}
  \label{eq:gc_pot_th_short}
  \Omega_{\rm th} = \Omega_{\rm q\bar{q}} + \Omega_{\rm B\bar{B}} +
  \Omega_{\rm M}.
\end{align}
In explicit form, the thermal energy contributions are defined as
\begin{align}
  \label{eq:gc_pot_thermal_qq}
    \Omega_{\rm q\bar{q}} = &- T \sum\limits_{i \in q}
    \frac{\gamma_i}{(2 \pi)^3} \int d^3k \; \left( \ln \left[ 1 +
        \Phi\, e^{ -\frac{1}{T} \left( E^*_i(k) - \mu^*_i
          \right)}\right] \right.\nonumber \\
      &\left. +\ln \left[ 1 + \bar{\Phi}\, e^{ -\frac{1}{T} \left(
              E^*_i(k) + \mu^*_i \right)} \right] \right),
\end{align}
for the quarks and antiquarks ($i \in \left\{ u,d,s \right\}$), which
couple to the effective Polyakov loop potential $\Phi$ and its
conjugate $\bar{\Phi}$ as described below,
\begin{align}
  \label{eq:gc_pot_thermal_BB}
  \Omega_{\rm B\bar{B}} = &- T \sum\limits_{j\in B} \frac{\gamma_j}{(2
    \pi)^3} \int d^3k \; \left( \ln \left[ 1 + e^{ -\frac{1}{T} \left(
          E^*_j(k) - \mu^*_j \right)} \right]\right.\nonumber \\
    &\hspace{5em}\left. +\ln \left[ 1 + e^{ -\frac{1}{T} \left(
          E^*_j(k) + \mu^*_j \right)} \right] \right),
\end{align}
for baryons and antibaryons with the sum running over all baryons $B$,
and
\begin{align}
  \label{eq:gc_pot_thermal_M}  
  \Omega_{\rm M} = &  T \sum\limits_{l\in M} \frac{\gamma_l}{(2
    \pi)^3} \int d^3k \; \ln \left[ 1 - e^{ -\frac{1}{T} \left(
          E^*_l(k) - \mu_l\right)} \right].
\end{align}
for the mesons, where in this case the sum runs over all mesons. The
spin-isospin-degeneracy factor is denoted with $\gamma_{i,j,l}$ for
the respective particle species $i,j,l$ and $E^*_{i,j,l}(k) =
\sqrt{k^2 + m_{i,j,l}^{*2}}$ for the single particle energies.\par
The introduction of the quarks in our model is done as in other recent
extensions of the original Nambu–Jona-Lasinio models~\cite{Nambu1961,
  Nambu1961b} which include an effective Polyakov loop potential. The
Refs.~\cite{Fukushima:2003fw, Meisinger:1995kp, Meisinger:1995ih,
  Ratti:2005jh, Ratti:2006gh, Ratti:2006wg, Roessner:2006xn} give a
comprehensive review of these PNJL models.\par
The scalar Polyakov loop field $\Phi$ in
Eq.~\eqref{eq:gc_pot_thermal_qq} is retrieved by tracing the constant
matrix valued temporal component of the SU(3) color gauge background
field
\begin{equation}
  \label{eq:polyakov-potential}
  \Phi = \frac{1}{3} \Tr{\left( e^{- A_0 / T } \right)}.
\end{equation}
Phenomenologically, this field indicates the phase transition from a
confined hadronic phase to a deconfined partonic phase and therefore
from the hadron resonance gas to the quark phase. These dynamics are
controlled by the effective Polyakov loop potential, adopted from
Ref.~\cite{Ratti:2006wg}, in the form
\begin{align}
  \label{eq:Polyakov-loop-eff-pot}
  \begin{split}
    U = -\frac{1}{2} \, a(T) \bar{\Phi}\Phi + b \left(
      \frac{T_0}{T} \right)^3 \ln \left[ 1 - 6
          \bar{\Phi}\Phi \right.\\
        \left. + 4 ( \bar{\Phi}^3 + \Phi^3 ) - 3 ( \bar{\Phi}\Phi )^2
        \right],
  \end{split}
\end{align}
which fulfills the Z(3) center symmetry of pure gauge QCD. The
temperature dependent parameter
\begin{align}
  \label{eq:Pol-loop-param}
  a(T) = a_0 + a_1 \left( \frac{T_0}{T} \right) + a_2 \left(
    \frac{T_0}{T} \right)^2
\end{align}
and the parameters therein are chosen such as to reproduce lattice
data for QCD thermodynamics in the pure gauge
sector~\cite{Boyd:1996bx} as well as known features of the
deconfinement transition~\cite{Ratti:2006wg}. By doing so, the
potential reproduces the pressure, the energy density, and the entropy
density in the pure gluonic sector given by lattice QCD calculations.
At low temperatures, in the confined phase, $\Phi$ is zero and its
value rises with increasing temperatures. It tends to values $\Phi
\rightarrow 1$ above the critical Polyakov temperature $T_0$ and with
increasing chemical potentials. Generally, in PNJL models, the gluon
dynamics are reduced to the static background field as given by the
Polyakov loop and to point couplings. As C.~Ratti et al. point out in
Ref.~\cite{Ratti:2005jh}, due to these simplifications of the gluon
dynamics and the insufficient treatment of transverse gluons, which
are dominant degrees of freedom at very high temperatures, the PNJL
model is likely to deviate from lattice QCD thermodynamics in the
high-temperature region (see also~\cite{Meisinger:2003id}).\par
The model parameters for the Polyakov loop potential are listed in
Tab.~\ref{tab:polyakov-model-parameters}. For the free parameter of
the Polyakov loop $T_0$, a value of $235$~MeV is chosen when used in
the non-interacting scenario, in which all couplings are set to zero,
and $225$~MeV, when all particles in the model couple to the
respective meson fields as described above.\par
\begin{table}[tb]
  \centering
  \begin{tabular}{|c|c|c|c|c|}
    \hline
    $T_0$ [MeV] & $a_0$ & $a_1$ & $a_2$ & $b$\\
    \hline
    225 -- 235 & 3.51 & -2.47 & 15.2 & -1.75 \\
    \hline
  \end{tabular}
  \caption{Parameters of the effective Polyakov loop potential
    Eqs.~\eqref{eq:Polyakov-loop-eff-pot}, \eqref{eq:Pol-loop-param}
    and coupling parameters of the model. The value of $T_0$ is
    adapted according to couplings (see text).} 
  \label{tab:polyakov-model-parameters}
\end{table}

Since the frozen dilaton field $\chi = \chi_0$ constrains the chiral
condensate to finite values for all $T$ and $\mu$, a coupling of the
Polyakov loop potential to the dilaton field is introduced. This
suppresses $\sigma$ when deconfinement is achieved. It is realized via
\begin{equation}
  \label{eq:coupling-polchi}
  \chi = \chi_0\left[ 1 - \left( \frac{\Phi^2}{2}  +
      \frac{ \bar{\Phi}^2 }{2}  \right)^2 \right], 
\end{equation}
a form similar to the one described in
Ref.~\cite{Steinheimer2011}. This coupling leads to a stronger and
more reasonable suppression of the chiral condensate in the presence
of free quarks in the last three terms of the scalar meson self
interaction in Eq.~\eqref{eq:self_int_scalar_mes}.\par
Minimizing the grand canonical potential
Eq.~\eqref{eq:grand_canon_pot} for a given temperature and
baryochemical potential with respect to its respective fields, leads
to a coupled system of non-linear equations of motion for the fields
and the particle densities $\rho_{i,j,l}$. In the following, all
thermodynamic quantities are derived starting with the pressure $p =
-\partial \Omega / \partial V$ and the entropy density $s = \partial p
/ \partial T $ by using the expression for the internal energy
$\epsilon = Ts - pV + \sum_i \mu_i \,\rho_i$, with the summation index
$i$ running over all particle species in the model.\par
In order to use hadronic degrees of freedom at low temperatures and
densities and quarks at higher temperatures and densities, we
introduce an eigenvolume $V_{\rm ex}^i$ for each particle species $i$,
as it was done in Refs.~\cite{Rischke1991, Mishra:2007cz,
  Steinheimer:2010ib}; similar approaches to repulsive volume
interactions are carried out in Refs.~\cite{Cleymans:1992jz,
  Ma:1993qz, Yen:1997rv, Gorenstein:1999ce, Bugaev:2000wz}.  By doing
so, it is ensured that with increasing densities at higher $T$ and
$\mu$ -- at the latest when quark abundances rise quickly at the
deconfinement phase transition -- an effective suppression of hadronic
degrees of freedom is established due to excluded volume effects, as
was shown in Refs.~\cite{Baacke1977, Hagedorn1980, Gorenstein1981,
  Hagedorn1983}.  An eigenvolume for every particle of a species $i$
is introduced and expressed by
\begin{equation}
  \label{eq:excluded_volume}
  V_{\rm ex}^i = \frac{1}{2}\; \frac{4}{3} \pi (2\,r)^3,
\end{equation}
with the parameter $r$ defining the effective radius of the hadrons in
the model. For the relation between the baryon and the meson radii, we
assume $r_M = r_B / 2$. For calculations including the quark phase and
assuming full couplings to the meson fields, a value $V_{\rm ex}^B =
2.7$~fm for the baryons and $V_{\rm ex}^B / 8$ for the mesons is
chosen (cf.\ Tab.~\ref{tab:excluded-vol-baryons}). Quarks do not
exhibit a finite volume.\par
In this way, each single hadron occupies a certain volume and thereby
reduces the total volume $V$ by $V_{\rm ex}^i$.  This results in a
modified chemical potential
\begin{equation}
  \label{eq:chem_pot_new}
  \mu_j^* = \mu_j - \sum \limits_i V_{\rm ex}^{i} p^i,
\end{equation}
where $p^i$ denotes the partial pressure of species $i$.\par
The particle, energy, and entropy densities are corrected by the
excluded volume and found to be
\begin{equation}
  \label{eq:corrected_densities}
  \rho_i' = \rho_i \xi_i^V,\qquad e' =    \sum\limits_i
  e_i \xi_i^V, \qquad s' =   \sum\limits_i s_i  \xi_i^V ,
\end{equation}
with the correction factor given as $\xi_i^V = 1/(1 + V_{\rm ex}^i
\rho_i)$. This definition of $\mu_j^*$ and the volume corrected
thermodynamic quantities ensures that thermodynamic consistency holds
true in the whole $T$-$\mu$-plane. That means, the relation for the
internal energy, as given above, and all partial derivatives for
determining the thermodynamic quantities remain valid.

\section{Results}
\label{sec:results}

\subsection{Model properties}
\label{sec:model_properties}

\begin{figure}[tb]
  \centering
  \includegraphics[width=.6\columnwidth]{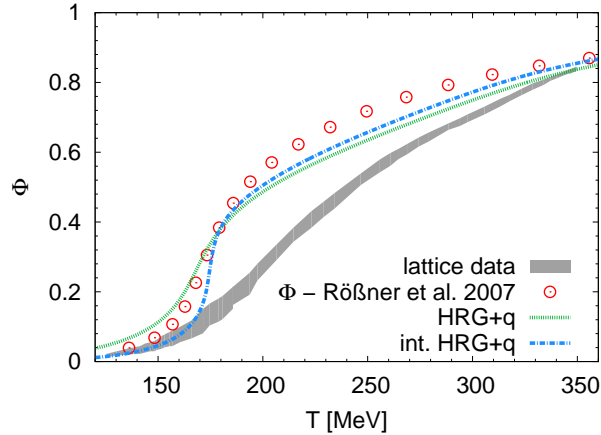}
  \caption{(Color online) The Polyakov loop order parameter $\Phi$ for
    the deconfinement phase transition as a function of $T$ at $\mu =
    0$. For the two scenarios including the quark phase, the
    non-interacting scenario (green line) and the interacting scenario
    (blue line), $\Phi$ only shows rather small differences below the
    critical temperature $T_c \approx 170$~MeV. Our results are close
    to those from Ref.~\cite{Roessner:2006xn} (red points) from which
    the Polyakov loop potential was adopted. Other parameterizations of
    the Polyakov loop potential (e.g.\ given in
    Ref.~\cite{Fukushima:2008wg}) yield a very similar slope of
    $\phi$. The gray band covers the lattice QCD results for $\Phi$
    given in Refs.~\cite{Bazavov2009a, Bazavov2010e, Bazavov:2010bx,
      Cheng2009, Cheng2010, Borsanyi2010, Borsanyi2010b, Aoki2009}.}
  \label{fig:pol_mu0}
\end{figure}

First, we present results from the described model at vanishing
baryochemical potential, $\mu_B = 0$.
\begin{figure}[tb]
  \centering
  \includegraphics[width=.6\columnwidth]{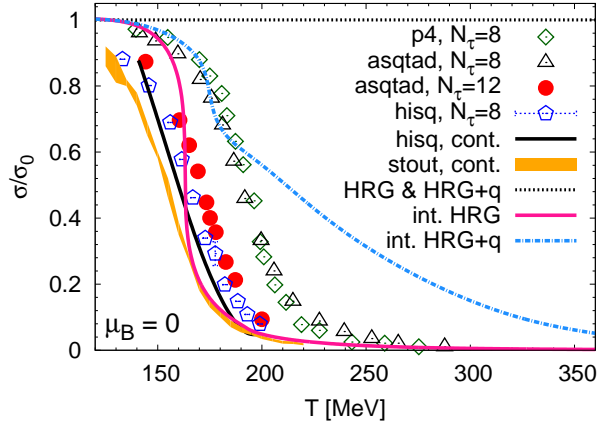}
  \caption{(Color online) Normalized order parameter for the chiral
    condensate $\sigma/\sigma_0$ as a function of the temperature
    compared to lattice data from Refs.~\cite{Bazavov2009a,
      Bazavov2010e, Bazavov:2010bx, Cheng2009, Cheng2010,
      Borsanyi2010, Borsanyi2010b, Aoki2009}. The red solid line
    depicts the results from the model for the interacting hadron
    resonance gas (HRG) without a quark phase. The interacting HRG
    with a quark phase is shown by the blue dashed line and the green
    dashed line shows the results for hadrons that do not interact
    with the mesons fields, together with the described quark
    phase. Comparing the curves, it clearly shows that even when
    including quarks, the chiral phase transition is mainly driven by
    hadronic degrees of freedom.}
  \label{fig:order_param_mu0}
\end{figure}
Figure~\ref{fig:order_param_mu0} shows the order parameter for the
chiral phase transition normalized to its ground state value
$\sigma/\sigma_0$ as a function of the temperature. The order
parameter for the deconfinement transition $\Phi$ as a function $T$ is
shown in Fig.~\ref{fig:pol_mu0}. The results from our model are
compared to lattice QCD results (data points or gray band) using
different improved staggered fermion actions. In detail, these are the
asqtad action from Ref.~\cite{Bazavov2009a}, the HISQ action from
Ref.~\cite{Bazavov2010e,Bazavov:2010bx,Bazavov:2011nk}, the p4 action
from Ref.~\cite{Cheng2009,Bazavov2009a,Cheng2010}, and the stout
action from Ref.~\cite{Borsanyi2010,Borsanyi2010b,Aoki2009}. For the
calculations with the described chiral model, here and in the
following, four different scenarios, i.e.\ basic parameter sets with
the following abbreviations, are distinguished:
\begin{description}
\item[HRG] describes the pure hadron resonance gas without excluded
  volume suppression and a quark phase that does not interact with the
  mesonic fields.
\item[Int. HRG] same as above with the only difference, that all
  hadrons couple to the meson fields as described above.
\item[HRG+q] includes the hadrons with excluded volume effects and a
  quark phase. All couplings to the mesonic field are set to zero only
  the quarks couple to the Polyakov loop.
\item[Int. HRG+q]. Same as above but here, all couplings to the meson
  fields are set to the finite values given in
  Tab.~\ref{tab:coupl-parameters}.
\end{description}
\begin{table}[tb]
  \centering
  \begin{tabular}{|c|c|c|c|c|c|}
    \hline
    & $V_{\rm ex}$  &  $g_{\rm Bs}$ &  $g_{\rm Bv}$ &  $g_{\rm qs}$ &
    $g_{\rm qv}$\\ 
    \hline
    HRG        & 0.0   &\text{\sffamily X} &\text{\sffamily X}
    &\cellcolor[gray]{0.9} &\cellcolor[gray]{0.9} \\  
    int.\ HRG   & 0.0   & \checkmark& \checkmark
    &\cellcolor[gray]{0.9} &\cellcolor[gray]{0.9} \\  
    \hline
    HRG+q      & 1.8 &\text{\sffamily X} &\text{\sffamily X}
    &\text{\sffamily X} &\text{\sffamily X} \\ 
    int.\ HRG+q & 2.7 & \checkmark& \checkmark & \checkmark &
    \checkmark \\
    \hline
  \end{tabular}
  \caption{The four different scenarios considered here include
    the non-interacting HRG and the interacting HRG, both with and
    without a quark phase q. For all scenarios, the values for
    the excluded volume $V_{\rm ex}$ are given in fm$^3$. The
    couplings of the baryons $B$ and the quarks $q$ to the scalar and
    the vector meson fields ($g_{\rm Bs,v}$,  $g_{\rm qs,v}$) are
    either set to zero when marked with a cross, or have the finite value
    listed in Tab.~\ref{tab:coupl-parameters} if indicated by a
    check mark (see text). The first two scenarios are purely
    hadronic and do not include a quark phase.}  
  \label{tab:excluded-vol-baryons}
\end{table}
A summarizing table of the configurations for these scenarios is given
in Tab.~\ref{tab:excluded-vol-baryons}. For the full interacting
scenario, a value $V_{\rm ex} = 2.7$~fm$^3$ is chosen for the
baryons. This corresponds to a mean baryon radius of $r_B =0.86$~fm
which is slightly smaller than the measured proton charge radius $r_p
= 0.88$~fm~\cite{mohr_2010_2011}.\par
The results for the purely hadronic resonance gas that is interacting
with the meson fields as described above (int.\ HRG) is depicted by
the solid magenta line (color online). In this case, at the critical
temperature $T_c = 164$~MeV, which quantitatively agrees well with
recent lattice data, $\sigma$ falls off in a small temperature
range. As already pointed out in Ref.~\cite{Rau:2011av}, a slower
decrease of $\sigma$, as it is observed in the lattice data, can be
achieved by taking into account pionic self
interactions~\cite{Steinheimer2011a, Gerber1989, Mishra2008}.
Nevertheless, as predicted by all lattice QCD calculations, the
transition is a smooth cross-over for both order parameters. As found
in Ref.~\cite{Rau:2011av}, the inclusion of all known hadronic degrees
of freedom in a $\sigma$-$\omega$ model restricts the chiral phase
transition to a smooth cross-over in the whole $T$-$\mu$
plane. Nevertheless, the first-order liquid-gas phase transition for
nuclear matter at the ground state up to a critical end-point $T_c
\approx 15$~MeV stays present in our model.\par
The second scenario considered here, is the interacting HRG with the
described PNJL-like quark phase (blue dashed line). The decrease of
$\sigma$ is also very strong but the slope flattens significantly at
the critical temperature for the Polyakov loop $T_c = 175$~MeV as
shown in Fig.~\ref{fig:pol_mu0}. However, in comparison to the pure
HRG it shows, that the slope of $\sigma(T)$ does not change up to
$T_c$ and therefore that the chiral phase transition is mainly driven
by hadronic degrees of freedom.\par
Since quarks in the model do not have an eigenvolume, quark degrees of
freedom are strongly preferred to be occupied at $T_c$ and above. As
of this temperature, the number of quarks exceeds the number of
baryons and mesons in the systems. Their smaller coupling strength to
the chiral condensate (cf.\ Tab.~\ref{tab:coupl-parameters}) leads to
a much slower decrease in $\sigma$. On the other hand, the rapid
increase of the quark multiplicities at $T_c$ generates a distinctly
steeper slope of the Polyakov loop parameter $\Phi$ (b) than predicted
by lattice QCD data, which is compiled in the gray band in
Fig.~\ref{fig:pol_mu0}.\par
In contrast to this finding, the Polyakov loop $\Phi$ results
reproduce those from Ref.~\cite{Roessner:2006xn} depicted by the red
circles. More recent PNJL models~\cite{Fukushima:2008wg} also show a
significant overshooting of the lattice results in the transition
region revealing an obvious discrepancy between effective models and
first-principle calculations on this point. In lattice QCD, there is
no sharp increase in the deconfinement order parameter at $T_c$.\par
As two limiting cases, the order parameters for the non-interacting
HRG is shown with [green dashed line in Fig.~\ref{fig:pol_mu0}] and
without a quark phase. The $\sigma$-parameter for the latter two
scenarios are shown by the flat line at $\sigma/\sigma_0 = 1$. The
excluded volume parameters of the baryons $V_{\rm ex}$ for the four
different scenarios considered had to be adapted in order to both give
a smooth transition between the two phases and a reasonable agreement
with lattice data for the thermodynamic quantities. These requirements
cause a rather large eigenvolume of the baryons in the fully
interacting scenario.\par
The values chosen for the different model scenarios considered in this
study are listed in Tab.~\ref{tab:excluded-vol-baryons}. In the two
scenarios with only a hadronic phase, on the other hand, hadrons do
not carry an explicit eigenvolume and thus excluded volume corrections
do not apply.\par
\begin{figure}[tb]
  \centering
  \includegraphics[width=.55\columnwidth]{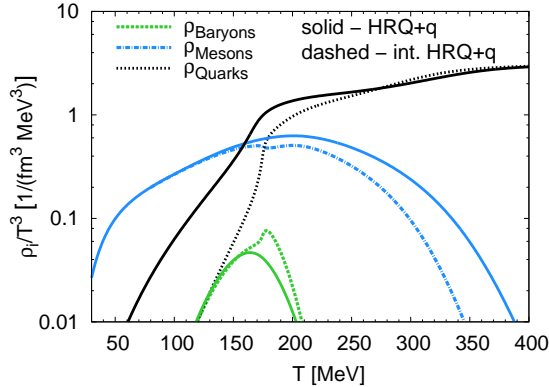}
  \caption{(Color online) Net baryon number densities over the
    temperature cubed $\rho_i/T^3$ of all baryons (green lines),
    mesons (blue lines) and quarks (black lines) as a function of the
    temperature at $\mu_B = 0$. Solid lines depict results for the
    ideal HRG and dashed lines the interacting HRG both including the
    quark phase.  For the eigenvolume of the baryons, $V_{\rm ex} =
    2.7$~fm$^3$ is chosen for the interacting case and $V_{\rm ex} =
    1.8$~fm$^3$ for the non-interacting HRG (see text). As expected,
    since there is no repulsive vector interaction for the quarks in
    the ideal scenario, the quark density below $T_c$ is much higher
    than in the fully interacting scenario.}
  \label{fig:particle_densities_mu0}
\end{figure}
Figure~\ref{fig:particle_densities_mu0} shows the total baryon number
densities of baryons and antibaryons (green lines), mesons (blue
lines), and quarks plus antiquarks (black lines) at $\mu = 0$ as a
function of the temperature, for the two scenarios containing a quark
phase. The results are shown for the non-interacting ideal gas
scenario (solid lines) and the fully interacting scenario with the
non-zero couplings to the fields (dashed lines). Obviously, the
coupling of the particles to the scalar fields, which generates the
effective masses as defined by Eq.~\eqref{eq:effective_mass}, leads to
an increase in the particle multiplicities at $T_c$, in particular of
the baryon multiplicity which, however, drops again shortly due to the
large baryon volume in favor of a quickly rising quark
multiplicity. In the temperature range $0.6 \, T_c < T < 1.3 \, T_c$,
there is a phase in which quarks and hadrons coexist, as expected for
a cross-over transition. At a temperature $T \approx 1.3 \, T_c$,
baryons vanish from the system.\par
In contrast, the quarks, show a stronger dependence on their couplings
to the repulsive vector fields. If the quarks do not couple to the
repulsive vector field, showing an unphysical behavior they appear
even at temperatures far below $T_c$ and their total density is much
larger than the baryon density for all temperatures. Obviously, this
behavior has a strong impact on the nuclear ground state what will be
studied below. If the mesons couple to the fields, the effective
masses of the pseudoscalar mesons scale as $m_{\rm mes}^{*2} \sim
1/\sigma$~\cite{Rau:2011av}. For this reason, at temperatures $T >
T_c$ their multiplicities decrease significantly faster than in the
non-interacting case due to an increasing mass.\par
This finding of hadrons even at temperatures above $T_c$, agrees well
with considerations that light and strange resonances can exist in
this region and are likely to be produced even in the QGP due to their
short formation times~\cite{Markert:2008jc, Bellwied:2010pr}. In
Ref.~\cite{Ratti:2011au} it was shown by comparing PNJL and lattice
QCD calculations that hadronic bound states are expected to exist up
to temperatures of $2\, T_c$.\par
As defined by Eq.~\eqref{eq:effective_mass}, the effective masses
$m^*$ of the particles in the chiral model are dynamically generated
by the coupling to the scalar meson fields. Increasing temperature and
density decreases $\sigma$ and $m^*$ accordingly. In consequence,
chiral symmetry can be restored.
\begin{figure}[tb]
  \centering
  \includegraphics[width=.5\columnwidth]{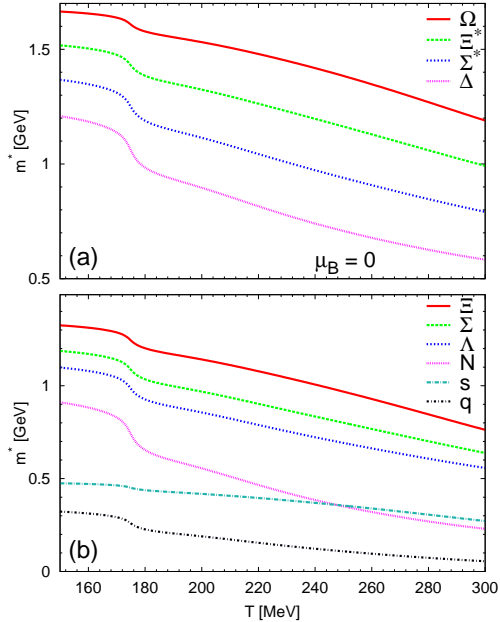}
  \caption{(Color online) Effective masses $m^*$ of the quarks (u,d =
    q and s), the baryonic octet (lower panel), and the decuplet
    (upper panel) for the fully interacting scenario including hadrons
    and quarks as a function of the temperature at $\mu_B = 0$. All
    shown particles couple to the fields with coupling strengths as
    listed in Tab.~\ref{tab:coupl-parameters}. The change of the
    effective masses largely reflects the slope of the scalar
    $\sigma$-field (cf.\ blue line in Fig.~\ref{fig:order_param_mu0});
    therefore the values for $m^*$ show a rather smooth decrease with
    increasing temperature.}
  \label{fig:mass_shift_mu0}
\end{figure}
Figure~\ref{fig:mass_shift_mu0} shows the effective masses of the
baryons and of the quarks in the fully interacting scenario as a
function of the temperature. The lower panel (b) shows the effective
masses of the baryon octet ($N$, $\Lambda$, $\Sigma$, $\Xi$), the up
and down valence quarks, denoted as $q$, and the strange quark s. In
the upper panel (a) the effective masses of the lowest lying baryon
resonances in the decuplet ($\Delta$, $\Sigma^*$, $\Xi^*$, $\Omega$)
are shown. Since the decrease of the scalar fields happens rather
smoothly with increasing temperature, the decrease of $m^*$ is also
relatively slow. Nevertheless the steepest decline is still visible at
$T_c$~\cite{Theis:1984qc, Waldhauser:1987uk}. For nucleons, the mass
at $T = 1.5 \, T_c$ has dropped down to $0.33 \, m_0^N$, for the
constituent up and down quarks to $0.27 \, m_0^{u,d}$ , and to $0.69
\, m_0^s$ for the strange quarks. Most baryons vanish at $T = 1.3\,
T_c$. For this reason, the mean field approximation should not be
affected by too small baryonic masses and possible fluctuations as
argued in Refs.~\cite{Mocsy2004, Bowman2009}.\par
In the hadronic version of this effective model, i.e.\ without the
quark phase and excluded volume correction
terms~\cite{Papazoglou:1997uw, Papazoglou:1998vr}, the couplings of
the baryon octet to the mesonic fields and the mesonic potential were
chosen in such a way as to reproduce the well-known vacuum masses of
the baryons and the experimentally determined masses of the mesons, as
well as the nuclear ground state properties (e.g.\ the correct binding
energy per nucleon and the nuclear compression modulus), and the
asymmetry energy. The implementation of the quark phase is done in a
way which ensures that these requirements are still
fulfilled. Moreover, one has to consider whether quarks already appear
in the nuclear ground state.\par
First, we consider temperature zero.
\begin{figure}[tb] 
  \centering
  \includegraphics[width=.55\columnwidth]{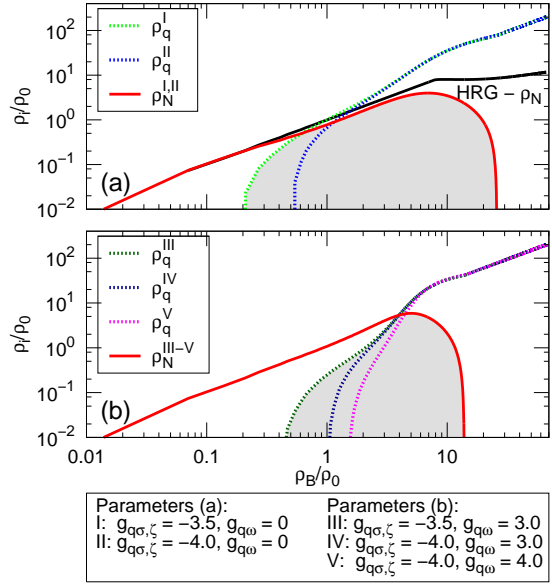}
  \caption{(Color online) Particle vector densities of up and down
    quarks $\rho_q$ (dashed lines) and of nucleons $\rho_N$ (solid red
    line) at $T = 0$ as a function of $\rho_B/\rho_0$ for zero (a) and
    non-zero quark vector couplings $g_{q\omega}$ (b) and different
    values of the scalar couplings $g_{q\sigma, \zeta}$ (see parameter
    list below the plots). The gray shaded areas indicate possible
    coexistence of quarks and nucleons. If the quark vector couplings
    take a finite value (b), quarks are predominantly present at
    $\rho_B = 4 \, \rho_0$ and all nucleons vanish within the model at
    $\rho_B = 15 \, \rho_0$. The solid black line in (a) depicts
    $\rho_N$ for the purely hadronic model without quarks. To ensure
    that no free quarks are present in the ground state, both
    couplings have to be chosen sufficiently large.}
  \label{fig:T0_densities}
\end{figure}
Figure~\ref{fig:T0_densities} shows the vector particle densities of
up and down quarks
\begin{equation}
  \rho_q = \sum\limits_{i = u,d} \rho^+_{q_i} - \rho^-_{\bar{q}_i}
  \label{eq:vector-densities}
\end{equation}
together with the density of the nucleons $\rho_N$ at $T = 0$, where
the anti-particle density $\rho^-_j$ is zero, as a function of the
baryonic density $\rho_B$ for different sets of the quark coupling
parameters (I--V, see key in figure) in the fully interacting model
including hadrons and quarks.  All densities are given in units of the
ground state density $\rho_0$. In panel (a) the densities for a
vanishing coupling of the quarks to the vector meson fields $g_{q
  \omega} = 0$ and for two different values for the quark - scalar
meson field coupling ($g_{q \sigma, \zeta} = -3.5$, blue doted line,
and $g_{q \sigma, \zeta} = -4.0$, magenta dashed line) are shown. It
is obvious that without the coupling to the repulsive vector field, in
both cases the quarks appear at densities smaller than the nuclear
ground state density at $\rho_B = 0.15$~fm$^3$. This specific choice
of quark coupling parameters may thus only be physically correct if
the quarks in this region of the phase diagram are not considered to
be unbound particles. The green dashed line in both panels of the
figure shows the density of the nucleons $\rho_N$ which differs only
slightly for the different parameter sets in (a) and (b). As a
reference, the solid black line in panel (a) depicts the nucleon
density in the purely hadronic model without a quark phase.\par
The values for the quark scalar coupling in panel (b) are the same as
before and the value of the up and down quark vector coupling is
varied between $g_{q \omega} = 3.0$ and $g_{q \omega} = 4.0$. It
shows, that with a scalar coupling of $g_{q \sigma} = -4.0$ for the up
and down quarks as well as $g_{q \zeta} = -4.0$ for the strange quarks
and a vector coupling equal or larger than $g_{q \omega} = 3.0$ for up
and down quarks, all quarks are effectively suppressed below the
nuclear ground state density as expected. Due to this finding, with
the given choice of parameters, which is compatible with the
corresponding parameters in the baryon sector, the quarks do not
affect the nuclear ground state. In the case of non-zero quark vector
couplings (b), quarks are predominantly abundant at $\rho_B = 4 \,
\rho_0$ and all nucleons within the model cease to exist at $\rho_B =
15 \, \rho_0$.\par
With this parameter set, the model reproduces the nuclear saturation
density $\rho_0 \approx 0.15$~fm$^3$ and the correct binding energy
per nucleon $E/A(\rho_0) \approx -16$~MeV. At the saturation density
the compression modulus of infinite nuclear matter is given
by~\cite{Blaizot:1980tw}
\begin{equation}
  \label{eq:compression-modulus}
  K_{\infty} = k^2_F \left. \frac{d^2\, E/A}{d\, k^2_F}
  \right|_{k_F=k_{F_0}} = 9 \rho^2_0 \left. \frac{d^2\, E/A}{d\,
      \rho^2} \right|_{\rho = \rho_0}.
\end{equation}
Calculations of isoscalar giant monopole and dipole resonances, i.e.\
compression modes of nuclei, and related experimental data suggest a
value of $ K_{\infty} = 240 \pm 20$~MeV for the compression modulus of
nuclear matter (see Ref.~\cite{Shlomo2006} for a recent comprehensive
review on how to determine $K_{\infty}$). The purely hadronic model
neglecting the eigenvolume of the particles, gives a value of
$K_{\infty} = 225$~MeV. However, if quarks are included and the
excluded volume effects are taken into account in order to suppress
hadronic degrees of freedom above $T_c$, the pressure and $E/A$ as a
function of the baryon density increases much faster depending on the
choice of $V_{\rm ex}$. This leads to a significantly higher
compressibility modulus, which is $K_{\infty} = 370$~MeV for an
eigenvolume of the baryons of $V_{\rm ex} = 1$~fm$^3$ and an
unreasonably high value of $K_{\infty} = 890$~MeV if $V_{\rm ex}$ is
set to $2$~fm$^3$.\par
This rapid increase in the density could be overcome by making the
eigenvolume of the particles temperature or density
dependent. Phenomenologically, it seems plausible that the eigenvolume
of one hadron is not perceived by another particle further away at low
temperatures and densities below the nuclear ground state, where
particles are statistically distributed relatively far from each other
in a large volume. For this reason, the eigenvolume effect would
vanish in this region and it would reach its actual full value at
higher densities, at which the particles in the system are densely
packed and would start to overlap. This, on the one hand, would still
result in the same effective suppression of hadrons at high
temperatures and densities in favor for point-like quarks as
before. On the other hand, the introduction of a density or
temperature dependent eigenvolume needs to be done with care, as not
to violate the thermodynamic consistency of the approach (cf.\
Eq.~\eqref{eq:corrected_densities})~\cite{Bugaev:2000wz,
  Kostyuk:2000nx, Bugaev:2008zz}. Particularly, since the transition
between the hadronic and the quark phase is abrupt in the fully
interacting model, this seems reasonable in order to achieve a much
smoother transition. Work along this line is in progress.

\subsection{Thermodynamics}
\label{sec:thermodynamics}

We will now take a closer look at the thermodynamic properties of the
model.
\begin{figure}[tb]
  \centering
  \includegraphics[width=.55\columnwidth]{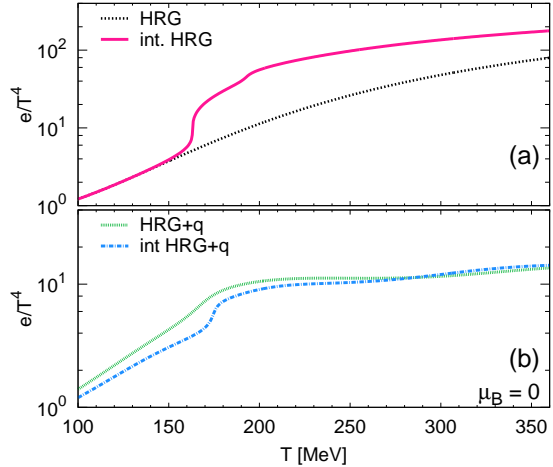}
  \caption{(Color online) Energy density over temperature to the
    fourth power at zero baryochemical potential from the presented
    model. Panel (a) shows results from the model without
    quarks. While the non-interacting HRG (black dashed line) rises
    monotonously, in the interacting scenario (solid magenta line)
    $e/T^4$ rises quickly at $T_c$ due to the drop in the hadron
    masses. Panel (b) shows the same quantity for the model including
    a quark phase. In the case without the meson field interactions
    (HRG+q -- green line), $e/T^4$ grows much faster as in the
    interacting case (HRG+q -- blue line), due to the non-existent
    vector repulsion of quarks. Unlike in a Hagedorn resonance gas,
    all quantities in the model saturate at very large temperatures
    caused by a large but still limited number of degrees of freedom.}
  \label{fig:pure_energy}
\end{figure}
Figure~\ref{fig:pure_energy} shows the energy density divided by the
temperature to the fourth power, $e/T^4$. Panel (a) depicts the purely
hadronic model without quarks in the non-interacting scenario (black
dashed line) as well as results from the scenario in which hadrons
fully interact with the meson fields (solid magenta line). The
interacting HRG shows a rapid rise of the energy at $T_c$. This is due
to the drop in the scalar condensate $\sigma$. Therefore, also the
effective hadron masses decrease rapidly, which leads to large hadron
abundances in the absence of any volume based particle repulsion. The
increase of the thermodynamic quantities of the ideal HRG is obviously
much slower above $T_c$ since the masses are not dynamically generated
via couplings to the scalar meson fields. The purely hadronic model is
considered only to be valid at least up to temperatures in the region
of $T_c$.\par
Panel (b) shows the results from the chiral model including the
additional quark phase. Below the critical temperature, the energy
density in the non-interacting scenario (green dashed line) is
significantly higher than in the non-interacting scenario (blue dashed
line). This is because in this case, there is no suppression by the
vector meson fields and quarks are not suppressed at low temperatures
(cf.\ Fig.~\ref{fig:particle_densities_mu0}). Furthermore, in this
scenario, the excluded volume of the hadrons $V_{\rm ex}$ is chosen a
bit smaller (cf.\ Tab.~\ref{tab:excluded-vol-baryons}). In the region
between $T = 140$~MeV and $170$~MeV the interacting scenario shows a
dip in the energy density. This is caused by the eigenvolume
suppression of the hadrons and a rather slow increase of the quark
abundances due to their coupling on the repulsive vector meson
fields. At higher temperatures ($T > 250$~MeV), results from both
model scenarios converge again, reflecting the fact that from this
temperature on the system in both scenarios is dominated by
quarks. Unlike in a Hagedorn resonance gas~\cite{Hagedorn:1968jf},
given the large but still limited number of degrees of freedom in our
model, thermodynamic quantities do not diverge.
\begin{figure}[tb]
  \centering
  \includegraphics[width=.55\columnwidth]{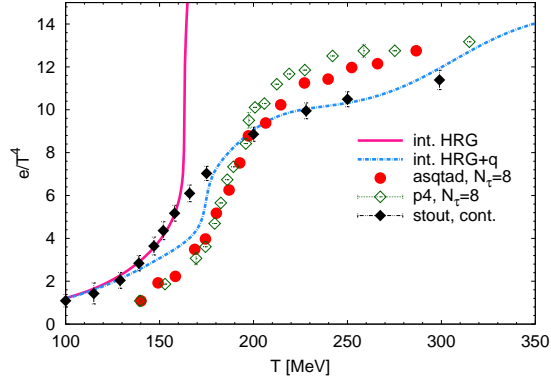}
  \caption{(Color online) Same quantity as above, again for the
    interacting HRG with (blue line) and without a quark phase
    (magenta line). The results are compared to lattice QCD
    data~\cite{Cheng2009, Bazavov2009a, Cheng2010, Borsanyi2010,
      Borsanyi2010b, Aoki2009}. Up to $T_c$, the newer stout continuum
    extrapolated data (black points) is in reasonable agreement to the
    pure interacting HRG. Compared to the pure HRG, the scenario
    including quarks shows a suppression of $e/T^4$ in the region of
    $T_c$. This is caused by a slower increase of the energy due to
    the eigenvolume suppression of hadrons and cannot be compensated
    by the quarks which just start to populate the system at these
    temperatures. Therefore, the model with quarks lies in the region
    between newest stout action results and the somewhat older lattice
    data (asqtad and p4 action) which show large deviations from the
    HRG at $T < T_c$.}
  \label{fig:energy_comp_lattice}
\end{figure}\par
Figure~\ref{fig:energy_comp_lattice} shows the same quantity from the
interacting HRG with and without quarks in comparison to lattice QCD
results from Refs.~\cite{Cheng2009, Bazavov2009a, Cheng2010,
  Borsanyi2010, Borsanyi2010b, Aoki2009}. Up to $T_c$, the results
from the purely hadronic model are in quantitatively good agreement
with newest lattice data with the continuum extrapolated stout
action~\cite{Borsanyi2010, Borsanyi2010b, Aoki2009}. Compared to the
HRG, the somewhat older lattice data from the HotQCD
collaboration~\cite{Bazavov2009a, Cheng2009, Cheng2010} clearly
underestimates the energy density in the hadronic regime ($T <
T_c$). In this figure, the impact of the quark phase and the
associated excluded volume suppression on the thermodynamic properties
of the model becomes obvious. As mentioned above, there is a dip in
$e/T^4$ in the vicinity of $T_c$. In this region, the rather large
eigenvolume of the hadrons leads to a flattening of the baryon density
as a function of the temperature, while the quark density is only just
starting to rise sharply and cannot compensate the lower hadron
contribution (cf.\ dashed lines in
Fig.~\ref{fig:particle_densities_mu0}). In the transition region,
results from our model including quarks lie in the region between the
lattice data from the HotQCD collaboration and newest continuum
extrapolated Wuppertal-Budapest results. A convergence of more recent
and highly anticipated data from the HotQCD collaboration, probably
with physical pion masses and continuum extrapolation, towards our
full-model results seems possible.\par
\begin{figure}[tb]
  \centering
  \includegraphics[width=.55\columnwidth]{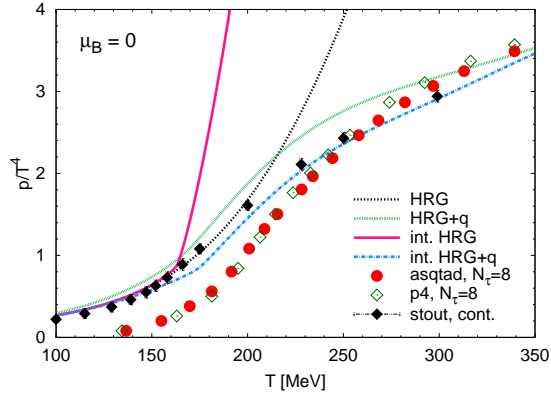}
  \caption{(Color online) Pressure over temperature to the fourth
    power from the chiral model in comparison to lattice QCD results
    at $\mu = 0$. The line styles depicting the different model
    scenarios are as in Fig.~\ref{fig:pure_energy}. As for the energy,
    the lattice data show obvious discrepancies between the different
    lattice actions. The model with presumably correct degrees of
    freedom, the interacting HRG with the quark phase, shows
    reasonably good agreement to latest lattice data (stout, continuum
    extrapolated).}
  \label{fig:pressure_model}
\end{figure}
Figure~\ref{fig:pressure_model} shows the pressure over the
temperature to the fourth power as a function of $T$. Depicted are
results from the presented model (with line styles as in the previous
figures) compared to lattice data. Here, the same observations as for
the energy can be made. The pressure in both scenarios without quarks
rises monotonously and again much faster above $T_c$ in the
interacting case due to the dropping effective masses. The scenarios
including quarks differ again, mainly due to their difference in the
coupling of particles to the vector fields. The pressure calculated
with the interacting model reproduces the slope of the stout action
results nicely, only undershooting lattice results in a small region
around $T_c$. Again, compared to HRG results, older lattice data
obviously yields too small values for $T < 1.2\, T_c$.\par
\begin{figure}[tb]
  \centering
  \includegraphics[width=.55\columnwidth]{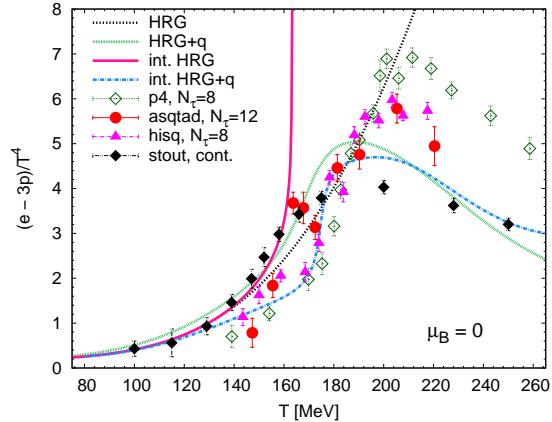}
  \caption{(Color online) Interaction measure, $(e - 3\,p)/T^4$, as a
    function of the temperature at $\mu_B = 0$ in comparison to
    lattice data obtained with different fermion actions. Line styles
    and colors for the different scenarios as depicted in
    Fig.~\ref{fig:pure_energy}, additional lattice data taken from
    Ref.~\cite{Bazavov2010e,Bazavov:2010bx}. Up to $T \approx T_c$ the
    data obtained with the continuum extrapolated stout action agrees
    well with the HRG with quarks scenario.}
  \label{fig:int_measure}
\end{figure}

An important quantity, characterizing the phase transition, is the
trace of the energy momentum tensor $e - 3\,p$, also known as
interaction measure or trace anomaly, the basic thermodynamic quantity
in lattice QCD from which other thermodynamic quantities can be
obtained via integration.  Figure~\ref{fig:int_measure} shows this
quantity (again divided by $T^4$) as a function of the temperature
calculated with the chiral model using the four different scenarios
described above.
As before, the results of the scenarios, which only include a hadronic
phase without a volume suppression effect, show a rapid rise at
$T_c$. The results of the other two scenarios, however, show a maximum
slightly above $T_c$. Again, the ideal non-interacting scenario
(dashed green line) shows a smoother transition between the hadronic
phase and the quark phase. The fully interacting model (dashed blue
line) exhibits a slowly increasing interaction measure below $T_c$,
comparable to the dip in the energy density in this region, followed
by a rapid rise due to dropping hadron masses and a slower decrease
above the critical temperature.\par
Results from the latter two scenarios show a slope comparable to those
from lattice QCD. Certainly, the chiral model does not reproduce the
results of one lattice action consistently. In the vicinity of $T_c$,
results from the ideal hadron and quark scenario (dashed green line)
yield a qualitatively good agreement with lattice results using the
continuum extrapolated stout action. With increasing temperatures ($T
> T_c$) all lattice data show a smooth decline towards small
values. This slope of the interaction measure is not reproduced with
the chiral model, which, contrastingly, shows a rather flat behavior
followed by a delayed decline at significantly higher temperatures ($T
\sim 350$~MeV). Since the applied Polyakov potential
Eq.~\eqref{eq:Polyakov-loop-eff-pot} has been fitted to reproduce
lattice results in the pure gauge sector~\cite{Ratti:2006wg}, this
deviation in the slope must be attributed both, to the slowly
decreasing meson fields in the presence of quarks and to hadrons
occurring at temperatures well above $T_c$ (cf.\
Fig.~\ref{fig:order_param_mu0} and
\ref{fig:particle_densities_mu0}). Additionally, at large
temperatures, transverse gluons are known to contribute largely to the
total pressure. However, the Polyakov loop corresponds exclusively to
the longitudinal gauge field and, therefore, underestimates the
pressure at very high $T$. In Ref.~\cite{Fukushima:2008wg}
K.~Fukushima shows that the applicability of the Polyakov potential
used in this work~\cite{Ratti:2006wg} breaks down above $T \sim 2\,
T_c$. This shortcoming could be another reason for the significant
deviation of model results from lattice QCD in the high-temperature
region.\par
A quantity which connects thermodynamic properties of the equation of
state with the dynamics of heavy ion collisions is the velocity of
sound $c_s$ which is deduced from thermodynamic quantities via
\begin{equation}
  \label{eq:sound_velocity}
  c_s^2 = \left. \frac{\partial p}{\partial \varepsilon}
  \right|_{s/\rho_B}.
\end{equation}
The square of the speed of sound of nuclear matter equals the change
in pressure for a given change in energy along an isentropic expansion
path. Thus, this important property controls the expansion dynamics of
the whole fireball, also of sound-like fluctuations and density
perturbations within the whole expanding system. The expansion
dynamics and possible high density shock waves evolving from the
collision of the two nuclei, in particular in uneven sized colliding
nuclei, are likely to be reflected in final state particle
observables~\cite{Rischke:1995mt, Dumitru:1998es, Bass:2000ib,
  Knoll:2011zz}. Vice versa, the determination of the speed of sound
from easily accessible particle observables may provide in-depth
information on the thermodynamic properties of nuclear matter,
especially in relation to the phase transition where $c_s$
significantly drops~\cite{Scheid:1974zz, Mohanty:2003va,
  Bravina:2008ra, Castorina:2009de, Rau:2010qy, Senger:2011zza,
  Steinheimer:2012bp}.\par
\begin{figure}[tb]
  \centering
  \includegraphics[width=.55\columnwidth]{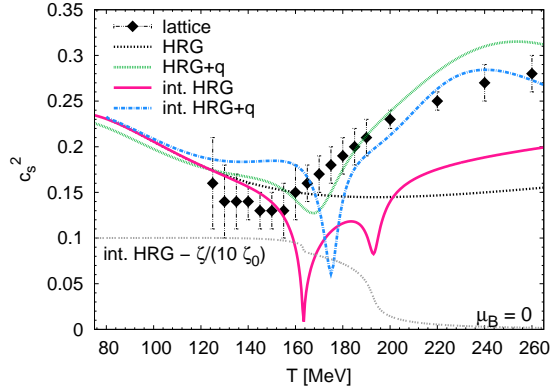}
  \caption{(Color online) Sound velocity squared at zero baryochemical
    potential for the different scenarios as described in the caption
    of Fig.~\ref{fig:pure_energy} as a function of the
    temperature. For both, the pure HRG and the HRG with quarks, the
    resulting equation of state softens considerably if the
    interactions of the particles with the mesonic fields are taken
    into account. The second dip for the interacting HRG at $T \approx
    190$~MeV, which is also slightly visible in the interacting HRG
    with the quark phase, is caused by a sharp drop of the
    $\zeta$-field (dashed gray line). The lattice data (stout,
    continuum extrapolated) is almost flat up to $T_c$ and and rises
    with a slope comparable to the HRG+q scenarios for higher $T$
    (taken from Ref.~\cite{Borsanyi:2012cr}).}
  \label{fig:sound_vel}
\end{figure}

The velocity of sound squared as a function of the temperature
calculated with the chiral model and from lattice
QCD~\cite{Borsanyi:2012cr} is shown in Fig.~\ref{fig:sound_vel}.
Again, the different parameter scenarios described above are
considered and, as expected, it is observed that the smoother the
transition of the thermodynamic quantities at $T_c$, the stiffer the
equation of state. This seems plausible since a smooth change in
effective degrees of freedom at the phase transition leads to much
smoother course in energy density and pressure and thus in a much
flatter derivative Eq.~\eqref{eq:sound_velocity}.\par
The interacting HRG (solid magenta line) shows the smallest $c_s^2$ at
$T_c$ due to the rapid increase of degrees of freedom in a small range
caused by the drop in $m^*$.  When looking at the speed of sound for
the two fully interacting scenarios, one also observes a second drop
down to $c_s^2 = 0.1$ (or a small dip in $c_s^2$ if the quark phase is
taken into account) at temperatures slightly above $T_c$. This local
minimum is caused by the strange quark-antiquark condensate $\zeta$
which falls off at this point and affects $m^*$ as described in
Eq.~\eqref{eq:effective_mass}. For a better understanding, the
$\zeta$-field ($0.1\, \zeta(T)/\zeta_0$) in the interacting HRG
scenario is illustrated by the thin dashed gray line. The ideal HRG
with the quark phase (green dashed line) shows only a small negative
deviation from the ideal HRG (black dashed line) at $T_c$. As
mentioned above, this is due to smooth changes in $p$ and $e$ at the
phase transition (cf.\ Fig.~\ref{fig:int_measure}). For higher
temperatures the value quickly tends toward the ideal gas limit $c_s^2
= 1/3$, as expected.\par
In experiment, a distinctive softening of the EoS, i.e.\ a small
$c_s^2$ in the transition region, strongly slows down the flow of
matter during the collision process and thus stops a deflection of
spectator matter. This, so-called, {\em burning log}
effect~\cite{Rischke:1995cm, Hung:1997du}, causes an observable
reduction of directed and elliptic flow of particles produced in
semi-peripheral collisions, as was argued in Refs.~\cite{Soff:1999yg,
  Li:1999gw, Zheng:1999gt, Stoecker:2004qu, Li:2005gfa}. This behavior
was experimentally observed by the NA49 collaboration
Ref.~\cite{Alt:2003ab}. At beam energies in the region between $30$
and $40$A~GeV, a collapse of the proton directed and elliptic flow was
found. In addition, as studied in Refs.~\cite{Rischke:1996em,
  Li:2008qm}, the softening of the equation of state in the transition
region should be reflected in a strictly non-monotonic behavior of the
HBT radii extracted from collisions in this intermediate energy
region.\par
In contrast to these findings, there is no softening of the EoS in the
lattice data. The sound velocity from Ref.~\cite{Borsanyi:2012cr}
stays constant up to $T_c$ and rises for higher temperatures with a
slope comparable to chiral model results with included quark
phase. This, in fact, would rule out a significant delay in the
expansion of the fireball and a prolonged lifetime of matter with
temperature below $T_c$ as suggested in Refs.~\cite{Rischke:1995cm,
  Rischke:1996em}.\par
Given these observations, we conclude that our EoS based on the
interacting HRG including the quark phase, offers not only feasible
degrees of freedom in a wide range of temperatures and densities but
also exhibits properties consistent with experimental observations.

\section{Summary}
\label{sec:summary}

In this paper, results from a well tested chiral effective
$\sigma$-$\omega$ model were presented, including all known hadronic
degrees of freedom, which for this study was extended by a quark
phase. The quarks were added in an PNJL-like approach in which they
couple to an effective Polyakov loop potential. The chiral order
parameter and the Polyakov loop order parameter for the deconfinement
transition show a critical temperature $T \approx 165$~MeV, which is
in line with all latest lattice QCD predictions. It shows, that the
chiral phase transition is mainly driven in the hadronic regime and
there are no qualitative changes concerning the chiral transition when
adding a quark phase.\par
At higher temperatures and densities hadronic degrees of freedom are
effectively suppressed in favor of a purely quark-dominated QGP phase
by eigenvolume corrections. Due to this fact, the model presented here
offers a description of nuclear matter with the correct degrees of
freedom in a wide range of temperature and baryochemical
potential. Additionally, with the chiral model it is possible to study
both, the chiral and the deconfinement phase transition. Fixing the
value for the coupling strength of quarks to the repulsive meson
vector fields, which has a major influence on the structure and the
properties of the resulting phase diagram, reasonable nuclear ground
state properties and a first-order liquid-gas phase transition can be
reproduced. Nevertheless, the value of the nuclear matter compression
modulus may restrict our choice of values for the finite particle
volume and the repulsive quark vector couplings. For a future study,
we intend to introduce a thermodynamically consistent implementation
of temperature or density dependent eigenvolume corrections that could
remedy this problem. For both, the chiral and the deconfinement phase
transition, a smooth cross-over transition at vanishing and also
finite baryochemical potentials is obtained, as previously shown
without quarks for the chiral transition in
Ref.~\cite{Rau:2011av}.\par
The thermodynamic quantities from the chiral model, generally show a
qualitatively reasonable agreement with lattice data, even though, in
particular for the interaction measure there are still significant
differences between the various lattice QCD results depending on the
lattice action used. The softening of the sound velocity in the
transition region was present in all scenarios except the pure HRG,
where no rapid change of degrees of freedom takes places. For the HRG
with the included quark phase the largest drop in $c_s^2$ was
observed. If all particles in the model fully couple to the mesonic
fields, the sound velocity squared gets close to zero at the phase
transition. This significant softening meets a major constraint on the
EoS as it is suggested by experimental observations on the drop in the
flow by the NA49 collaboration~\cite{Alt:2003ab}.\par
The chiral effective model presented here, is able to provide an EoS
to be used in hydrodynamic models for studying the dynamics of heavy
ion collisions. This is a major benefit of an integrated model which
has the right degrees of freedom in a wide range of the phase
diagram. Consequently, the resulting EoS is well suited for
calculations with hybrid models~\cite{Bass:2000ib, Teaney:2001av,
  Hirano:2005xf, Nonaka:2006yn, Song:2010mg, Yan:2011tn}, like the
UrQMD hybrid model~\cite{Petersen:2008dd}, since it exhibits known
features of QCD and has exactly the same degrees of freedom in the
hydrodynamic stage as in the early and subsequent cascade stage.

\section{Acknowledgements}
\label{sec:ack}

This work was supported by BMBF, GSI, and by the Hessian excellence
initiative LOEWE (Landesoffensive zur Entwicklung
Wissenschaftlich-\"okonomischer Exzellenz) through the Helmholtz
International Center for FAIR (HIC for FAIR), and the Helmholtz
Graduate School for Hadron and Ion Research (HGS-HIRe). Computational
resources were provided by the Center for the Scientific Computing
(CSC) of the Goethe University Frankfurt. J.~S.\ acknowledges a Feodor
Lynen fellowship of the Alexander von Humboldt foundation. The authors
thank M.\ Bleicher for fruitful discussion.

\bibliographystyle{iopart-num.bst}
\bibliography{Bib}

\end{document}